\documentclass[twocolumn,pra,aps,showpacs,floatfix]{revtex4}
\usepackage{epsfig}
\begin{document}
\title{One-Qubit and Two-Qubit Codes in Noisy State Transfer}
\begin{abstract}
Quantum state transfer is a procedure, which allows to exchange quantum information between stationary qubit systems. It is anticipated that the transfer will find applications in solid-state quantum computing. In this contribution, we discuss the effects of various, physically relevant models of decoherence on a toy model of six qubit linearly coupled by the exchange interaction. In many cases we observe the advantage of the two-qubit encoding, which can be associated with the fact that this encoding does not require the state initialization.
\end{abstract}
\author{Marcin Markiewicz}\affiliation{Institute of Theoretical Physics and Astrophysics, University of Gda\'nsk, PL-80-952 Gda\'nsk} 
\author{Marcin Wie\'sniak}\affiliation{Institute of Theoretical Physics and Astrophysics, University of Gda\'nsk, PL-80-952 Gda\'nsk}
\date{28.04.2009}
\maketitle
\section{Introduction}
Quantum computation, with algorithms such as those presented in \cite{DJ,Shor,Grover}, are one of the biggest promises of the quantum information science. In many different qubit implementations, the only way to exchange quantum information between elements of a complex system (e.g. a solid-state quantum computer) is through the short-range interaction between individual qubit units. This procedure is commonly known as the quantum state transfer \cite{bose}. 

Recent years have brought a significant interest in the development and better understanding of the state transfer mechanism. In most systems, the quantum message (the state to be tranferred) irreversibly spreads over the quantum wire (a chain of spins used for the state transfer). In principle, it is not possible to recover the whole information deterministically. Initial studies, e.g., \cite{danielbose,bgb,poznan}, aimed to counteract the effects of the non-periodic evolution. Thus authors have suggested measurement-based \cite{danielbose} and asymptotic \cite{bgb,poznan} protocols for reading the information. The significant progress was made together with the discovery of systems performing the perfect transfer \cite{christ,niko1,niko2} based on state mirroring \cite{shi}. %More recent articles describe diagnostics of a quantum wire with limited access \cite{danielkoji,paternostro1,paternostro2,marw}. 
Existence \cite{cubes} or impossibility \cite{pra} of state mirroring was proven for various lattices.

Likewise, a lot of attention was paid to the problem of the state transfer in presence of various types of decoherence \cite{aa1,aa2,aa3,aa4}. In this communication we aim to discuss the behavior of two different types of encoding. The fundamental difference in their robustness against various types of external noise, depends on whether or not the code involves coherences between states of a different total magnetization. 

The paper is organized in the following manner. Section II describes the Hamiltonian used for state transfer and the encodings discussed in this paper. In section III we compare the robustness of the encodings against some noise models. Thereafter comes the summary.
\section{System and Codes}
The standard system used for state transfer is a chain of spin-$\frac{1}{2}$ coupled by the exchange (xx) interaction of modulated strength,
\begin{equation}
\label{Hxx}
H_{xx}=\sum_{i=1}^{N-1}J_i(X^{[i]}X^{[i+1]}+Y^{[i]}Y^{[i+1]})+\sum_{i=1}^NB_iZ^{[i]}.
\end{equation}    
Here we adopt the following notation:
\begin{eqnarray}
X=\frac{1}{2}\left(\begin{array}{cc} 0&1\\ 1&0\end{array}\right),&Y=\frac{1}{2}\left(\begin{array}{cc} 0&-i\\ i&0\end{array}\right),\nonumber\\
Z=\frac{1}{2}\left(\begin{array}{cc} 1&0\\ 0&-1\end{array}\right).
\end{eqnarray}
The upper square-bracketed index denotes the spin being referred to by the operator. $J_i$ is the coupling strength and $B_i$ is the local magnetic field.

In general, these systems do not realize perfect state transfer. There is, however, a class of systems suitable for this task. They are symmetric with respect to the middle of the chain, and satisfy the spectrum parity-matching condition, formulated by Shi {\em et al.} \cite{shi} as:
\begin{eqnarray}
\label{spmc}
E_i=E_0+En_i,&& \Pi(|\psi_i\rangle)=(-1)^{n_0+n_i}.
\end{eqnarray}
Here, $|\psi_i\rangle$ denotes the $i$th eigenstate of the Hamiltonian with the corresponding energy $E_i$, $E_0$ and $E$ are some energetic constants, $n_0$ is some integer, $n_i$ is an integer function of $i$, and $\Pi(\cdot)$ is the parity function, yielding $+1$ for (spatially) even arguments and $-1$ for odd ones. As a consequence of this feature, at half of period $\tau$ the odd component of the global state picks a relative phase $\pi$ with respect to the even component. Thus initial state of the left side is mirrored to right and vice-versa. In particular, a system having this property was introduced in \cite{christ,niko1,niko2} and satisfies
\begin{eqnarray}
\label{christ}
J_i\propto\sqrt{i(N-i)},&&B_i=B.
\end{eqnarray}
This system will be considered throughout the paper. unless specified otherwise, the global magnetic field $B$ is chosen to be 0. The phase difference induced by the transfer between components of a different magnetization is fixed manually. 

As we have mentioned in the introduction, we aim to compare two different encodings for state transfer. One of them (a), called the one qubit code, is simply based on uploading the quantum message onto the first spin of the chain (the one closest to the sender). Logical $|0\rangle$ is represented by the maximally magnetized state of the system, $|00...0\rangle$, while to encode $|1\rangle$ we flip the first spin. Thus the initial operators coding the logical qubit, according to the Bloch formula,
\begin{eqnarray}
\rho=I+r_xX+r_yY+r_zZ\\
\left(I=\frac{1}{2}\left(\begin{array}{cc}1&0\\0&1\end{array}\right)\right)\nonumber,
\end{eqnarray} 
 are simply 
\begin{eqnarray}
X_{(a)}(t=0)=X^{[1]},&&Y_{(a)}(t=0)=Y^{[1]},\nonumber\\
Z_{(a)}(t=0)=Z^{[1]}.
\end{eqnarray}

The other method, the two-qubit code (b) strongly relies on state mirroring (satisfaction of (\ref{spmc})). First, we also initialize the chain in $|00...0\rangle$. To encode one of the qubit levels, we flip the first spin, to encode the other -- the second. Thus the quantum message is translated into two non-maximally magnetized states, $|10...0\rangle$ and $|01...0\rangle$. 
It is initially described by the following operators:
\begin{eqnarray}
\label{bcode}
&X_{(b)}(t=0)=X^{[1]}X^{[2]}+Y^{[1]}Y^{[2]},\nonumber\\&Y_{(b)}(t=0)=Y^{[1]}X^{[2]}-X^{[1]}Y^{[2]},\nonumber\\
&Z_{(b)}(t=0)=\frac{1}{2}\left(Z^{[1]}-Z^{[2]}\right).
\end{eqnarray}

Only because the initial state is perfectly reflected at certain times, the two initial coding states are locally orthogonal at the receiver's end of the chain. This is never the case when the evolution is not periodic.

We need to make a remark that the unit matrix in this encoding is given by $I_{(b)}(t=0)=I^{[1]}I^{[2]}-Z^{[1]}Z^{[2]}$. The receiver has a read-out head, which projects the final state onto the subspace of projector $P=2(I^{[N-1]}I^{[N]}-Z^{[N-1]}Z^{[N]})$. When this projection fails, the received state is assumed to be maximally mixed. Hence, the effective state in this protocol reads $\rho_{(b),eff}(\tau/2)=P\rho_{(b)}(\tau/2)P+(1-\text{Tr}P\rho_{(b)}(\tau/2)P)I_{(b)}(\tau/2)$, where $\tau$ is the period of the evolution (in the one excitation subspace (OES)).

In a recent contribution \cite{marw} we have shown that when encoding (b) given by (\ref{bcode}) is used, one does not need to initialize the state as $|00...0\rangle$, but rather just to make sure that the two first sites are initially factored out from the rest of the chain. State initialization is taken here only for the sake of a fair comparison. 
\section{Models of Noise}
Let us compare the robustness of these two encodings against various models of decoherence of potential physical interest. This will be done mainly by a numerical integration of the Lindblad equation of motion \cite{lindblad},
\begin{eqnarray}
\label{lind}		
\frac{\partial \rho}{\partial t}=&\frac{1}{i}[H,\rho]\nonumber\\
+&\sum_i\gamma_i\left(K_i\rho K_i^\dagger-\frac{1}{2}(K_i^\dagger K_i\rho+\rho K_i^\dagger K_i)\right).
\end{eqnarray}
$\gamma_i$'s are environment coupling strengths, and $K_i$'s are interaction operators. Our strategy throughout this paper is to integrate Eq. (\ref{lind})  numerically. While the free evolution is continuous, the interaction with the evironment is decomposed into at least 30 steps per transfer time. The calculations were performed for a toy model of a chain of 6 sites.

As a measure of the quality of the transfer we will take the fidelity averaged over all input states, 
\begin{eqnarray}
F&=&\frac{1}{4\pi}\int_{r_x^2+r_y^2+r_z^2=1}d\sigma\text{Tr}\nonumber\\
&(&\text{Tr}_{1,2,...,N-2(,N-1)}L(\rho\otimes|00...\rangle\langle 00...|)\nonumber\\
&\otimes&{\text{Tr}_{1,2,...,N-2(,N-1)}U\rho\otimes|00...\rangle\langle 00...| U^\dagger}),
\end{eqnarray} 
where $d\sigma$ is an element of the Bloch sphere, $\int_{r_x^2+r_y^2+r_z^2=1}d\sigma=4\pi$, $L(\rho\otimes|00...\rangle\langle 00...|)$ is the solution of Eqn. (\ref{lind}) at time $t=\tau/2$ and $U=\exp(i\tau H/2)$ is the unitary evolution realizing the perfect transfer. $\otimes|00...\rangle\langle 00...|$ stands for the fully magnetized state of this part of the chain, which does not contain the message at $t=0$. The partial trace excludes the last site for (a), and the last two spins for (b). This measure simplifies to
\begin{equation}
F=\frac{1}{2}+\frac{1}{6}(F^X+F^Y+F^Z),
\end{equation}
where  $F^A=\frac{1}{2}\text{Tr}(\text{Tr}_{1,2,...,N-2(,N-1)}L(A\otimes|00...\rangle\langle 00...|)(\text{Tr}_{1,2,...,N-2(,N-1)}UA\otimes|00...\rangle\langle 00...| U^\dagger) (A=X,Y,Z)$ is the ``fidelity'' of an operator transfer. The state transfer is meaningful and has quantum features only if the average fidelity is above $\frac{2}{3}$. Otherwise, the same amount of information could be sent by the means of classical communication. Basically, the sender could measure any component of his spin and announce the result. Then it would suffice for the receiver to align his spin accordingly. 

The first interesting model of decoherence is the depolarizing channel, in which all qubits are equally exposed to a random twirl,
\begin{eqnarray}
\label{twirl}
&\{K_i\}_{i=1}^{3N}=\{X^{[i]},Y^{[i]},Z^{[i]}\}_{i=1}^{N},\nonumber\\
&\gamma_i=\gamma.
\end{eqnarray}
This model is highly relevant to the present state-of-the-art ``qubit'' realizations. Currently, many systems used as qubits, e. g. Josephson junctions, Ryndberg atoms, or quantum dots are indeed high-dimensional systems \cite{alickiqubit}. There is a probability that such a system migrates from the qubit levels to higher ones, leaving the state of the effective qubit completely undetermined. In solid-state physics this effect could be, for example, due to interaction with phonons.

The result of the numerical simulation for are presented in Fig 1.
\begin{figure}[!t]
\centering
\includegraphics[width=4cm]{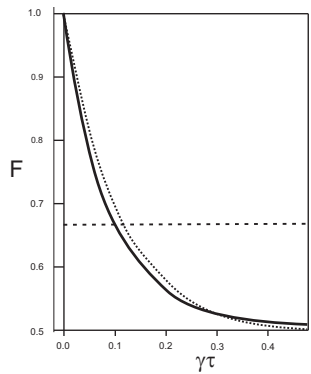}
\label{figtwirl}
\caption{The average fidelities of the state transfer in presence of a twirl on every qubit, Fidelities for the one- (solid line) and two-qubit codes (dotted line) are plotted in function of the coupling strength $\gamma$ (in units $1/\tau$). The dashed horizontal line is the classical threshold of $2/3$.}
\end{figure}
The comparison of the two codifications brings interesting observations. Initially, the two-qubit encoding is more efficient, but at some point code (a) becomes more robust against local dephasing. This, however, happens when the fidelity drops below the classical limit.

A recent result described in \cite{marw} provides a better understanding of this graph. One should keep in mind that it is not $X^{[1]}$ or $Y^{[1]}$, which evolves to $X^{[N]}$ or $Y^{[N]}$, but rather $N$-site operators $X^{[1]}Z^{[2]}Z^{[3]}...$ and $Y^{[1]}Z^{[2]}Z^{[3]}...$. State transfer with one-qubit encoding necessarily involves multi-site correlations. This is otherwise as in (b), where the initial  state of rest of the wire is irrelevant. It is then easy to see why this model of noise initially affects more (a). The error acts also on the part of the chain, which is essential for the transfer in the first scheme, even though it does not carry the message.  

This model was also studied for $N=7$. Again, we see the initial advantage of (b), which gradually decreases, and finally (a) provides more faithful transfer. The result is presented in Fig. 2.
\begin{figure}[!t]
\label{twirl7}

\centering
\includegraphics[width=4cm]{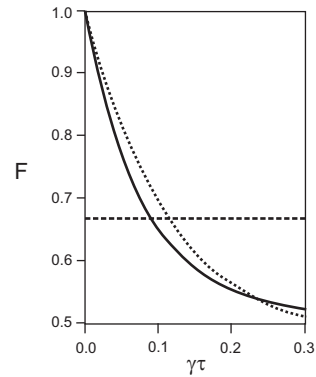}
\caption{The average fidelities $F^{(a)}$ (solid) and $F^{(b)}$ (dotted) in presence of a twirl on every site for $N=7$.}
\end{figure}
The crossover point is shifted towards lower $\gamma$ and higher fidelities, as compared with $N=6$. It remains an open and interesting question, where is the limit of this point in the regime of $N\rightarrow\infty$. Nevertheless, it seems natural to expect that for a weak coupling (b) overtakes (a), in accordance to the intuitive argument above. 
%One can also argue that the one qubit encoding is more robust in the weak coupling limit. This could be approximated by the situation in which decoherence acts only once on the system during the whole transfer (say, at time $t=0$),
%\begin{eqnarray}
%L(\rho_{in})&=&\rho_{in}\nonumber\\
%&+&\sum_i\gamma\left(K_i\rho_{in} K_i^\dagger-\frac{1}{2}(K_i^\dagger K_i\rho_{in}+\rho_{in} K_i^\dagger K_i)\right),&\\
%\rho_{out}&=&UL(\rho_{in})U^\dagger.
%\end{eqnarray}
%The effect of this map is rescaling all instances of the Pauli operators $X^{[i]},Y^{[i]},Z^{[i]}$ used to encode the state by the factor $1-4\gamma$. Thus we have $F_{(a)}(\gamma\tau\rightarrow 0)=1-2\gamma\tau$ and $F_{(b)}(\gamma\rightarrow 0)=1-4\gamma$. Hence the one qubit encoding is more efficient for the twirl decoherence.

The other potentially interesting from the physical point of view interaction with the environment is given by
\begin{eqnarray}
\label{localK}
&\{K_i\}_{i=1}^{2N}=\{S_+^{[i]},S_-^{[i]}\}_{i=1}^N,&\nonumber\\
&\gamma_{2i-1}=\gamma\exp(-\beta B),&\nonumber\\
&\gamma_{2i}=\gamma.
\end{eqnarray} 	 
This describes a situation, in which local states $|0\rangle$ and $|1\rangle$ are separated by energy $B$, which is large 
compared to couplings, and each qubit is coupled to its individual heat bath of inverse temperature $\beta\propto\frac{1}{T}$ (in natural units). $S_{\pm}$ stands for the local raising (lowering) operator, $X\pm iY$. 

The Hamiltonian of the whole system, including the interacting electromagnetic field, is given by
\begin{equation}
H_{total}=H_{XX}+H_{Int}+H_F,
\end{equation}
The field Hamiltonian given by $H_F=\sum_{\sigma=H,V}\int d^3k\omega a_{\vec{k},\sigma}^\dagger a_{\vec{k},\sigma}$, with $a_{\vec{k},\omega,\sigma}$ being the annihilation operator of a photon of momentum $\vec{k}$, frequency $\omega=\omega(\vec{k})$ and polarization $\sigma$. The interaction term is $H_{Int}=\sum_{j=1}^N \vec{D}^{[j]}\vec{E}_{reg}(\vec{r}_j)$. $\vec{D}^{[j]}=2\vec{d}X^{[j]}$ is the electric dipole operator ($\vec{d}$ is the dipole moment of the qubit), and 
\begin{eqnarray}
\vec{E}_{reg}(\vec{r})&=&i\sum_{\sigma=H,V}\int_{k<K_0} d^3k\left(\frac{2\pi|\vec{k}|}{L^3}\right)^{1/2}\nonumber\\
&\times&\vec{\epsilon}_{\vec{k},\sigma}(e^{i\vec{k}\vec{r}}a_{\vec{k},\sigma}+e^{-i\vec{k}\vec{r}}a_{\vec{k},\sigma}^\dagger).
\end{eqnarray}
$\epsilon_{\vec{k},\sigma}$ stands for the light polarization vector and $K_0$ denotes the formal cut-off, necessary for intermediate calculations, but to be put equal to infinity in the final result. 

As shown in \cite{AlickiLendi}, if the initial state of the field was thermal,
\begin{eqnarray}
\label{alendi}
\frac{\partial\rho}{\partial t}&=&-i[H,\rho]\nonumber\\
&+&\frac{1}2\sum_{m,n=1}^N a_{mn}\left(2S_-^{[m]}\rho S_+^{[n]}-S_+^{[n]}S_-^{m}\rho-\rho S_+^{[n]}S_-^{m}\right.\nonumber\\
&&+\left.e^{-\beta B}(2S_+^{[m]}\rho S_-^{[n]}-S_-^{[n]}S_+^{m}\rho-\rho S_-^{[n]}S_+^{m})\right).
\end{eqnarray}
In the regime of short waves, the matrix $\{a_{mn}\}$ becomes diagonal, and by the similarity between all spins is just proportional to the unit matrix. In this way we reach Eqns. (\ref{localK}). 

The results of calculation of the state transfer fidelity with this noise are presented in Fig. 3. The solid line denotes the threshold of $F_{(a)}=\frac{2}{3}$, the dashed one stands for $F_{(b)}=\frac{2}{3}$. The transfer with the codes has quantum features only in regions left to the respective lines, as the coupling constant $\gamma$ increases.

\begin{figure}[!t]
\label{figtherm}
\centering
\includegraphics[width=4cm]{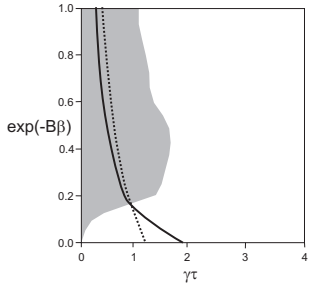}
\caption{The diagram for average fidelities of state transfer in presence of the interaction with short thermal electromagnetic waves. Left from the solid line $F^{(a)}>2/3$, the dotted line bounds the region of $F^{(b)}>2/3$ and in the grey region one has $F^{(b)}>F^{(a)}.$}
\end{figure}

As seen from Fig. 3, encoding (a) turns out to be more robust against this model of decoherence than (b) at low values of $\beta$, while it is less optimal at low temperatures. The result for high temperatures can be also explained in reference to \cite{marw}. At high temperatures, decoherence destroys correlations necessary for one-qubit operator transfer, as the steady state is the maximally mixed state. Transfer (b) does not rely on such correlations and thus is more robust. On the other hand, at low temperatures the steady state is the maximally magnetized state. Only the excitation carrying the message is affected. In scheme (a), $|00...\rangle$ is already part of the code, while encoding (b) takes place purely in OES. The OES operators is at $\beta=\infty$ decay in time like $e^{-\gamma t}$, being replaced by the vacuum projector and giving $F^{(b)}=\frac{1}{2}(1+e^{-\gamma\tau/2})$ The average fidelity in transfer (a) reads $F^{(a)}=\frac{1}{6}(3+e^{-\gamma\tau/2}+e^{-\gamma\tau/4})$

Let us also discuss the model of local fluctuating magnetic fields,
\begin{eqnarray}
\label{Kfluct}
&\{K_i\}_{i=1}^N=\{Z^{[i]}\}_{i=1}^N.\nonumber\\
&\gamma_i=\gamma
\end{eqnarray}
The results are presented in Fig. 4.
\begin{figure}[!t]
\centering
\label{figfluct}
\includegraphics[width=4cm]{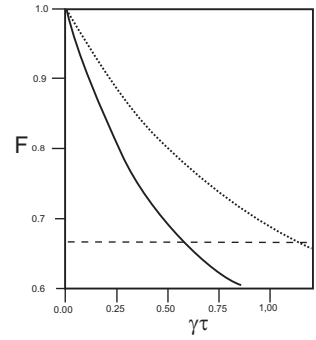}
\caption{The average fidelities of the state transfer in presence of a locally fluctuating magnetic field. The solid line represents $F^{(a)}$, the dotted-$F^{b}$.}
\end{figure}
This time, (b) turns out to be always more robust than (a). However, the nature of this result cannot be explained with \cite{marw} and remains unknown.
%The above results suggest that when the main source of decoherence are local perturbations, the one-qubit code is, in general, at least as robust to the coherence losses, as the two-qubit code. This is in agreement with our intuitive expectations. During the evolution, the two qubit encoding involves on average more correlations than the other encoding. Local dephasing models affect properties of all individual qubits; $i$-fold correlations are $i$ times more sensitive to distractions than localized properties.

Let us now discuss global models of decoherence. The non-local version of the twirl, in which 
\begin{equation}
\{K_i\}_{i=1}^3=\left\{\sum_{j=1}^NX^{[j]},\sum_{j=1}^NY^{[j]},\sum_{j=1}^NZ^{[j]}\right\}.
\end{equation}
will not be discussed, as it has weak associations with a physical model. Thus the next study will be conducted on the fluctuating global magnetic field,
\begin{equation}
\{K_i\}_{i=1}=\left\{\sum_{j=1}^NZ^{[j]}\right\}.
\end{equation}
In this case, the fidelity in the one-qubit encoding initially decays linearly with time, $F_{(a)}(\gamma\tau_1\rightarrow 0)= 1-2\gamma_1\tau$, while encoding $(b)$ turns out to be decoherence free. Operators, which are used to store the information at $t=0$, as well as the Hamiltonian, commute with the error operator $K_1$. Thus the message encoded into two qubits is not affected at all by the global fluctuations of the field.

This is not the case when the chain is exposed to the interaction with a long electromagnetic wave. 
In the regime of waves, the length of which (inverse proportional to $B$) is much longer than distances between spins \footnote{$B$ should be, however, dominant over $J_{i}$'s, so that the energy is almost completely determined by the magnetization.}, the matrix $\{a_{mn}\}$ in Eq. (\ref{alendi}) has all elements equal. Thus there only two $K_i$ operators,
\begin{eqnarray}
\{K_{i}\}_{i=1}^2&=&\left\{\sum_{j=1}^NS_-^{[j]},\sum_{j=1}^NS_+^{[j]}\right\},\nonumber\\
\{\gamma_i\}_{i=1}^2&=&\{\gamma,e^{-\beta B}\gamma\}
\end{eqnarray}
Figure 5 shows regions of $\gamma$ and $\beta B$, in which encodings (a) and (b) preserve no quantum features. 

\begin{figure}[!t]
\centering
\includegraphics[width=4cm]{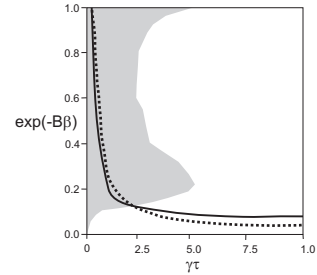}
\label{figlongwave}
\caption{The average fidelities of the state transfer in presence of interaction with long electromagnetic waves. The meaning of lines as in the caption of Fig. 3}
\end{figure}
The results and their general interpretation are similar to these with local thermal baths. A new element would be that at $\beta=\infty$ there exist states, which are not affected by this model of noise. $K_i$'s are the raising and the lowering operators of the total angular momentum, hence all singlet states are annihilated by their action. This brings out the following question: whether the state transfer could be made unaffected by the interaction with long waves if the message is encoded in two singlets. To make it possible, the partners should have granted access not to two sites at each side, but to four. The first non-trivial subspace of singlets is observed for this number of spins-$\frac{1}{2}$, and contains two states,
\begin{eqnarray}
|s_1\rangle&=&\frac{1}{2}(|01\rangle-|10\rangle)(|01\rangle-|10\rangle)\nonumber\\
|s_2\rangle&=&\frac{1}{\sqrt{3}}\big(|1100\rangle+|0011\rangle\nonumber\\
&-&\frac{1}{2}(|01\rangle+|10\rangle)(|01\rangle+|10\rangle)\big).
\end{eqnarray} 
However, it is easy to verify that also the message encoded in with these to states in the beggining of the wire is still exposed to the coherence loss. The reason for this is that only Heisenberg interactions without magnetic field preserve the total angular momentum, and this class of spin chains does not have the property of state mirroring \cite{pra}. Hence, even if the encoding does not involve any angular momentum, it will be generated during transfer, giving the opportunity for the environment to interact.

\section{Conclusions}

The numerical evidence presented in this contribution shows that for specific, physically relevant, noise models, such as the interaction with the fluctuating magnetic field or high-temperature electromagnetic waves, the two-qubit encoding provides more faithful transfer than the one-qubit code. This can be associated with the fact that codification (b) does not require the state initialization. When strategy (a) is adopted, the encoding has only an illusionary local character. In fact, the transfer is dependent on classical correlations of the rest of the chain. In genreral, however, the advantage of one scheme over the other depends on many conditions, such as the coupling strength, the particle number or temperature.

This allows to make a guess about the fidelities in the dual rail scheme \cite{danielbose}, when it is applied to $xx$ models. It is very similar to the second encoding, as the whole transfer takes place in the subspace of a fixed excitation number. The protocol involves, however, two independent systems, and each one of them must be initialized in a fully magnetized state. One should expect that whenever (a) is less optimal than (b) due to the necessity of the state initialization, it will be the same case for the dual-rail scheme. Also, this scheme employs the double infrastructure giving the environment more ways to interact with the system, thus supposingly accelerating the coherence decay in some cases (e.g., the twirl). Nevertheless, one should notice that the dual-rail encoding is insensitive to the global fluctuating magnetic field, just as (b) is.

An interesting question is if another initial state could improve the situation with encoding (a). The answer is negative. As we have mentioned the only operators, which are relevant for the transfer of $X^{[1]}$ and $Y^{[1]}$ are $X^{[1]}Z^{[2]}Z^{[3]}...$ and $Y^{[1]}Z^{[2]}Z^{[3]}...$. The fully magnetized state of the chain is chosen as the easiest state to be prepared, i. e. a product state, also being the ground state of the model for large $B$.

We have also argued that for the interaction with long magnetic waves there is no subspace, which allows to encode the message without exposing it to decoherence. This follows from the impossibility of the state mirroring for the angular momentum preserving systems.

We acknowledge R. Alicki, M. Horodecki, and A. Kay for useful remarks. The work is part of EU 6FP programmes QAP and SCALA and has been done at the National Centre for Quantum Information at Gda\'nsk.

\end{document}